
\input phyzzx
\voffset=0.25truein
\hsize=6truein
\def\TITLEPAGE{\frontpagetrue}
\def\CALT#1{\hbox to\hsize{\tenpoint \baselineskip=12pt
        \hfil\vtop{\hbox{\strut CALT-68-#1}
        \hbox{\strut DOE RESEARCH AND}
        \hbox{\strut DEVELOPMENT REPORT}}}}

\def\CALTECH{\smallskip
        \address{California Institute of Technology, Pasadena, CA 91125}}
\def\TITLE#1{\vskip 1in \centerline{\fourteenpoint #1}}
\def\AUTHOR#1{\vskip .5in \centerline{#1}}

\def\ABSTRACT#1{\vskip .5in \vfil \centerline{\twelvepoint \bf Abstract}
        #1 \vfil}
\def\ENDTITLEPAGE{\vfil\eject\pageno=1}

\def\sqr#1#2{{\vcenter{\hrule height.#2pt
      \hbox{\vrule width.#2pt height#1pt \kern#1pt
        \vrule width.#2pt}
      \hrule height.#2pt}}}

\def\section#1#2{
\noindent\hbox{\hbox{\bf #1}\hskip 10pt\vtop{\hsize=5in
\baselineskip=12pt \noindent \bf #2 \hfil}\hfil}
\medskip}

\def\underwig#1{        
        \setbox0=\hbox{\rm \strut}
        \hbox to 0pt{$#1$\hss} \lower \ht0 \hbox{\rm \char'176}}

\def\bunderwig#1{       
        \setbox0=\hbox{\rm \strut}
        \hbox to 1.5pt{$#1$\hss} \lower 12.8pt
         \hbox{\seventeenrm \char'176}\hbox to 2pt{\hfil}}

\def\MEMO#1#2#3#4#5{
\frontpagetrue
\centerline{\tencp INTEROFFICE MEMORANDUM}
\smallskip
\centerline{\bf CALIFORNIA INSTITUTE OF TECHNOLOGY}
\bigskip
\vtop{\tenpoint
\hbox to\hsize{\strut \hbox to .75in{\caps to:\hfil}\hbox to 3.8in{#1\hfil}
\quad\the\date\hfil}
\hbox to\hsize{\strut \hbox to.75in{\caps from:\hfil}\hbox to 3.5in{#2\hfil}
\hbox{{\caps ext-}#3\qquad{\caps m.c.\quad}#4}\hfil}
\hbox{\hbox to.75in{\caps subject:\hfil}\vtop{\parindent=0pt
\hsize=3.5in #5\hfil}}
\hbox{\strut\hfil}}}
\tolerance=10000
\hfuzz=5pt

\def\vslash{\rlap{/}v}

\TITLEPAGE
{\hbox to\hsize{\tenpoint \baselineskip=12pt
        \hfil\vtop{\hbox{\strut CALT-68-1783}
        \hbox{\strut UCSD/PTH 92-17}
        \hbox{\strut DOE RESEARCH AND}
        \hbox{\strut DEVELOPMENT REPORT}}}}
\TITLE {Baryons Containing a Heavy Quark as Solitons\foot{Work supported in
part by the U.S. Dept. of Energy under Contract no. DEAC-03-81ER40050 and under
grant no. DE-FG03-90ER40546, and by the National Science Foundation under a
Presidential Young Investigator
award no. PHY-8958081.}}
\medskip
\centerline{(REVISED VERSION)}
\AUTHOR{Elizabeth Jenkins and Aneesh V. Manohar}
\bigskip
\centerline{{\it Department of Physics, University of California, San Diego,}}
\centerline{{\it 9500 Gilman Drive, La Jolla, CA 92093-0319}}
\bigskip
\AUTHOR{Mark B. Wise}
\CALTECH
\ABSTRACT{The possibility of interpreting baryons containing a single
heavy quark as bound states of solitons (that arise in the nonlinear
sigma model) and heavy mesons is explored.  Particular attention is paid
to the parity of the bound states and to the role of heavy quark
symmetry.}

\ENDTITLEPAGE

\eject

Many properties of baryons $B$ containing light $u$ and $d$ quarks suggest
that they can be represented as solitons.  For example, they have a mass of
order $1/\alpha$ and the cross section for $e^+ e^- \rightarrow B \bar
B$ is of order $e^{-1/\alpha}$, where $\alpha = 1/N_c$ is taken as a
small quantity.$^1$  Skyrme originally suggested that baryons are solitons
in the chiral Lagrangian (i.e. nonlinear sigma model) used to describe
pion self interactions.$^2$  The solitons of the chiral Lagrangian have the
right quantum numbers to be QCD baryons,$^3$ provided one includes
the Wess-Zumino term.$^4$  The model of QCD baryons as
solitons has been used to compute many of their properties.$^{5,6}$  The same
large $N_c$ power counting that suggests that baryons containing only
light quarks are solitons, also suggests that baryons containing a single
heavy quark $Q$ ($m_Q \gg \Lambda_{QCD}$) can be represented as
solitons.

Recently there has been considerable progress in understanding the
properties of hadrons containing a single heavy quark.  In the limit
$m_Q \rightarrow \infty$, QCD has a heavy quark spin-flavor
symmetry$^{7,8}$ that
determines many properties of hadrons containing a single heavy quark.
In the $m_Q \rightarrow \infty$ limit, the total angular momentum of the
light degrees of freedom (i.e. light quarks and gluons),
$$     \vec S_\ell = \vec S - \vec S_Q \,\, , \eqno (1)$$
is conserved.  Consequently mesons and baryons containing a single heavy
quark can be labeled by $s_\ell$, and provided $s_\ell \not= 0$, they
come in degenerate doublets$^9$ with total spins
$$      s_\pm = s_\ell \pm 1/2  \eqno (2)$$
formed by combining the spin of the heavy quark with the angular
momentum of the light degrees of freedom.

Callan and Klebanov originally suggested an interpretation of baryons
containing a heavy quark as bound states of solitons of
the pion chiral
Lagrangian with mesons containing a heavy
quark.$^{10,11}$  In this paper we examine this model for heavy baryons.  An
important aspect of this work is that the consequences of heavy quark
symmetry are taken into account.  Heavy quark symmetry implies that a
doublet of mesons containing the heavy quark must be considered and that
both members of the doublet can occur in the bound state.  In this
regard the work presented here differs from that of Callan and Klebanov.
An important feature of the work of Callan and Klebanov was that the
ground state of the heavy baryon is in an $L = 1$ partial wave,
and thus the lowest mass
baryon containing a heavy quark has positive parity in accord with
experiment.  In this paper both the soliton and meson are taken as
infinitely heavy and it is assumed that the soliton-meson potential is
minimized at the origin.  Consequently, the spatial wavefunction for
the ground state is a Dirac delta function $\delta^3 (\vec x)$
and corresponds
to the lowest energy classical configuration which has the soliton  and heavy
meson located at the same spatial point.  We give a simple explanation of the
Callan-Klebanov parity flip appropriate to this description of the bound
state.

The lowest lying heavy mesons with $Q\bar q_a$ $(q_1 = u, q_2 = d)$
flavor quantum
numbers have $s_\ell = 1/2$ and come in a doublet containing spin-zero and
spin-one mesons.  For $Q = c$ these are the $D$ and $D^*$ mesons.  The
interactions of these heavy mesons with pions are described by a chiral
Lagrangian which is invariant under both heavy quark spin symmetry and
chiral $SU(2)_L \times SU(2)_R$ symmetry.\foot{We will restrict the analysis to
the case of two light flavors; the generalization to three flavors is
straightforward.}
Chiral $SU(2)_L\times SU(2)_R$ symmetry is spontaneously broken to the vector
$SU(2)_V$ subgroup, so the Goldstone boson manifold is
$SU(2)_L\times SU(2)_R/SU(2)_V$. A general Goldstone boson field configuration
is obtained by making a space-time dependent chiral transformation
$\left(g_L(x),g_R(x)\right)$ on a standard
vacuum state. The Goldstone boson fields $\Xi(x)$ in a
$G/H$ sigma model are defined by$^{12}$
$$
g(x) = \Xi(x) h(x) \eqno(3)
$$
where $g(x)\in G$ describes the locally rotated vacuum, $h(x)\in H$, and
$\Xi(x)$ is generated by $n_g$ independent broken generators, where $n_g$ is
the number of Goldstone bosons. Different choices for the $n_g$ broken
generators lead to different (but equivalent) nonlinear realizations
of the spontaneously broken chiral symmetry.
The standard $\Sigma$ basis for the QCD chiral
Lagrangian is defined by
$$
\left(g_L(x),g_R(x)\right) \equiv \left(\Sigma(x),1\right)\cdot
\left(V_\Sigma(x),V_\Sigma(x)\right) =
\left(\Sigma(x) V_\Sigma(x),V_\Sigma(x)\right), \eqno(4)
$$
where the broken generators are chosen to be the $SU(2)_L$ generators $T^A_L$.
The field $\Sigma(x)$ is equal to $g_L(x)g_R^\dagger(x)$, and
transforms under chiral $SU(2)_L \times SU(2)_R$ as
$$
\Sigma(x) \rightarrow L ~\Sigma(x) ~R^{\dagger} \,\, , \eqno (5)$$
since $g_L(x)\rightarrow Lg_L(x)$, $g_R(x)\rightarrow R g_R(x)$.
The pion fields occur in the $2
\times 2$ matrix $M$,
$$      M = \left[\matrix{\pi^0/\sqrt{2} & \pi^+\cr
\pi^- & -\pi^0/\sqrt{2}\cr} \right] \,\, , \eqno (6)$$
where
$$      \Sigma = \exp \left({2iM\over f}\right) \,\, , \eqno (7)$$
and $f$ is the pion decay constant.  Experimentally $f \simeq 132 \
{\rm MeV}$.  In
the large $N_c$ limit $f \sim 1/\sqrt{\alpha}$ (recall that $\alpha =
1/N_c$).
 For considering the couplings of heavy mesons to the pseudo-Goldstone
bosons, it is useful to introduce the $\xi$ basis defined by
$$
\left(g_L(x),g_R(x)\right) \equiv \left(\xi(x),\xi^\dagger(x)\right)
\cdot\left(V_\xi(x),V_\xi(x)\right) =
\left(\xi(x) V_\xi(x),\xi^\dagger(x) V_\xi(x)\right) \eqno(8)
$$
where the broken generators are chosen to be the axial generators
$T^A_L-T^A_R$.
Under chiral $SU(2)_L \times SU(2)_R$, $\xi$ transforms as
$$      \xi(x) \rightarrow L\ \xi(x)\ U^{\dagger}(x) =
U(x)\ \xi(x)\ R^{\dagger}\,\, ,\eqno (9)$$
where $U$ is a complicated function of $L,R$ and the pion fields and in
general depends on the space-time coordinates. However, for an unbroken
$SU(2)_V$
transformation, $V=L=R,$ $U(x)$ is equal to $V$, and is constant.
Comparing eq.~(8) with eq.~(4), we see that $\xi$ and $\Sigma$ are related
by
$$
\Sigma(x) = \xi^2(x). \eqno(10)
$$

The Goldstone boson manifold $SU(2)_L\times SU(2)_R/SU(2)_V$ is diffeomorphic
to the group manifold $SU(2)$, which is diffeomorphic to the three-sphere
$S^3$. The standard identification of $\Sigma$
with $S^3$ is obtained by writing
$$
\Sigma = a + i \vec b \cdot \vec \tau, \qquad a^2 +
|\vec b|^2 = 1, \eqno(11)
$$
and identifying $\Sigma$ with the point $(a,\vec b)\in S^3$.
The $\Sigma$ basis provides a single-valued coordinate system that
covers the entire surface of $SU(2)$, because
$$
\left(g_L,g_R\right)=\left(
\Sigma V_\Sigma, V_\Sigma\right)=\left(\Sigma' V'_\Sigma, V'_\Sigma
\right)\eqno(12)
$$
implies that $\Sigma=\Sigma'$.

Unlike the $\Sigma$ basis, the $\xi$ basis is not a well-defined
coordinate system on $SU(2)$. Two fields $\xi$ and $\xi'$ denote the same
Goldstone boson configuration if
$$
\left(g_L,g_R\right)=\left(\xi(x) V_\xi(x),\xi^\dagger(x) V_\xi(x)
\right) = \left(\xi'(x) V'_\xi(x),\xi'^\dagger(x)
V'_\xi(x)\right).
\eqno(13)
$$
This equality is satisfied if and only if
$$
\xi' = \xi h,\qquad h \xi h = \xi,\qquad h\in SU(2).\eqno(14)
$$
It is straightforward to solve this equation for $h\in SU(2)$. Denote the
general $SU(2)$ matrix $\xi$ by
$$
\xi = \alpha + i \vec \beta \cdot \vec \tau, \qquad \alpha^2 +
|\vec \beta|^2 = 1 .\eqno(15)
$$
If $\alpha\not=0$, then $\xi$ and $-\xi$ are equivalent. However, if
$\alpha=0$, all configurations of the form $i \vec \beta \cdot \vec \tau$ are
equivalent, irrespective of the value of $\vec\beta$. If the vector
$(\alpha,\vec\beta)$ is considered to be a point on the three sphere $S^3$,
then we see that points on the northern and southern hemisphere of $S^3$
obtained by inversion through the origin are equivalent, and the entire equator
is identified to a point. (The resulting manifold is still equivalent to
$S^3$.) Thus the $\xi$ basis gives a coordinate system for the Goldstone
bosons which is well-defined except near the equator of $S^3$, $\xi=i
\vec \beta \cdot \vec \tau$.
In the $\Sigma$ basis, the points on the entire three sphere $S^3$
correspond to inequivalent field configurations, and the $\Sigma$ basis
gives a well-defined coordinate system on all of $S^3$.

The chiral Lagrangian for heavy mesons is easy to construct in the
$\xi$ basis since the heavy meson
fields transform in a simple way under parity.
Heavy quark symmetry is incorporated into the chiral Lagrangian
by defining a heavy meson field which contains both pseudoscalar
and vector mesons.  The heavy meson field for the ground state
$Q \bar q_a$ mesons
is written as a $4 \times 4$ matrix$^{13}$
$$      H_a = {(1 + \vslash)\over 2} [P_{a\mu}^* \gamma^\mu - P_a
\gamma_5] \,\, , \eqno (16)$$
where $v^\mu$ is the heavy quark four-velocity, $v^2 = 1$.  We shall work
in the rest frame of the heavy mesons where $v^\mu = (1,\vec 0)$.  The
fields $P_a, P_{a\mu}^*$ destroy the heavy pseudoscalar and vector
particles that comprise the ground state $s_\ell = 1/2$ doublets.  The
vector meson field is constrained   by $v^\mu P_{a\mu}^* = 0$.  Under
$SU(2)_Q$ heavy quark spin symmetry
$$      H_a \rightarrow SH_a \,\, , \eqno (17)$$
where $S \in SU(2)_Q$, and under chiral $SU(2)_L \times SU(2)_R$
$$      H_a \rightarrow H_b U_{ba}^{\dagger} \,\, . \eqno (18)$$
It is also convenient to introduce
$$      \bar H_a = \gamma^0 H_a^{\dagger} \gamma^0 = [P_{a\mu}^{*\dagger}
\gamma^\mu + P_a^{\dagger} \gamma_5] {(1 + \vslash)\over 2} \,\, . \eqno
(19)$$
Under parity
$$      H (x^0, \vec x) \rightarrow \gamma^0 H(x^0, - \vec x) \gamma^0
\eqno (20)$$
and
$$      \xi (x^0, \vec x) \rightarrow \xi^{\dagger} (x^0, - \vec x)
\,\, . \eqno (21)$$
The chiral Lagrangian density for heavy meson-pion interactions
is$^{14,15,16}$
$$ {\cal L} = - i Tr \bar H_a v_\mu \partial^\mu H_a + {i\over 2} Tr
\bar H_a H_b v^\mu [\xi^{\dagger} \partial_\mu \xi + \xi \partial_\mu
\xi^{\dagger}]_{ba}$$
$$ + {ig\over 2} Tr \bar H_a H_b \gamma^\nu \gamma_5 [\xi^{\dagger}
\partial_\nu \xi - \xi \partial_\nu \xi^{\dagger}]_{ba} + ... \,\, ,
\eqno (22)$$
where the ellipsis denotes terms with more derivatives and repeated $a,b$
indices are summed over $1,2$.  Eq.~(22) is
the most general Lagrangian density
invariant under chiral $SU(2)_L \times SU(2)_R$,
heavy quark spin symmetry and parity.  It is easy to
generalize this Lagrangian density to include
explicit $SU(2)_L \times SU(2)_R$ symmetry breaking from $u$ and $d$
quark masses and explicit $SU(2)_Q$ symmetry breaking from
$\Lambda_{QCD}/m_Q$ effects.

The coupling $g$ determines the $D^* \rightarrow D\pi$ decay width,
$$      \Gamma (D^{*+} \rightarrow D^0 \pi^+) = {1\over 6\pi} ~{g^2\over
f^2} |\vec p_\pi|^3 \,\, . \eqno (23)$$
The present experimental limit,$^{17}$ $\Gamma (D^{*+} \rightarrow D^0
\pi^+)
\lsim\ 72\ {\rm keV}$, implies that $g^2  \lsim 0.4$.  In this work the sign of
$g$ plays an important role.  Applying the Noether procedure to the
Lagrangian density in eq.~(22) and to the QCD Lagrangian density gives
$$      \bar q_a \gamma_\mu \gamma_5 T_{ab} q_b = - g Tr \bar H_a H_b
\gamma_\mu \gamma_5 T_{ba} + ... \,\, , \eqno (24)$$
where $T$ is a traceless $2 \times 2$ matrix and the ellipsis denotes
terms containing the pion fields.  Treating the quark fields in eq.~(24)
as constituent quarks and using the nonrelativistic quark model to
evaluate the $D^*$ matrix element of the l.h.s. of eq.~(24) gives$^{16}$
$g = 1$.  (A similar estimate for the pion nucleon coupling gives $g_A =
5/3$.)  In the chiral quark model$^{18}$ there is a constituent-quark pion
coupling.  Using the chiral quark model gives $g \simeq 0.75$.

The soliton solution of the $SU(2)_L\times SU(2)_R$ chiral Lagrangian
for pion self interactions is
$$      \Sigma = A(t) \Sigma_0 (\vec x) A^{-1} (t) \,\, , \eqno(25)$$
where
$$      \Sigma_0 = \exp ~\left(i F(|\vec x|)\ \hat x \cdot \vec\tau\right)
\,\, ,
\eqno (26)$$
and $A(t)$ contains the dependence on collective coordinates associated
with rotations and isospin transformations of the soliton solution,
$$      A = a_0 + i \vec a \cdot \vec\tau \,\, , \eqno (27)$$
where $a_0^2 +| \vec a|^2 = 1$.  The soliton has the quantum numbers
of a baryon
containing light $u$ and $d$ quarks.
For solitons with baryon number one, the
function $F(|\vec x|)$ satisfies
$F(0) = -\pi$ and $F(\infty) = 0$.\foot{There appears to be a sign error in the
formula for the baryon number current used in Ref.~5.} Its detailed shape
depends on higher
derivative terms in the chiral Lagrangian for pion self interactions.
Since $F$ increases as $|\vec x|$ goes from zero to infinity, it is
expected that $F'(0)$ is positive.  The baryons containing $u$ and $d$
quarks have wavefunctions that are functions of the $a_\mu$.  For the
neutron and proton states; $|p\uparrow\rangle =
(1/\pi) (a_1 + ia_2), ~|p\downarrow
\rangle
= - (i/\pi) (a_0 - ia_3), {|n\uparrow\rangle} =
(i/\pi) (a_0 + ia_3) ~{\rm and}~
|n\downarrow
\rangle = - (1/\pi) (a_1 - ia_2) \,\, .$  The soliton $\Sigma_0(x)$ is a
winding
number one map from spacetime into the Goldstone boson manifold. In the $\xi$
basis, the soliton solution has the form (up to an overall sign)
$$
\xi_0 = \exp ~\left(i F(|\vec x|)\ \hat x \cdot \vec\tau/2\right).
\eqno(28)
$$
As $F$ varies from $F=0$ at
spatial infinity to $F=-\pi$ at the origin, $\xi_0$ varies
from $\xi_0=1$, the north pole of $S^3$, to $\xi_0=-i\vec\tau\cdot\hat
x$, the
equator, with intermediate points covering the northern hemisphere of $S^3$.
However, we have already seen that the Goldstone boson manifold in the $\xi$
basis is the northern hemisphere of $S^3$, with the equator identified to a
point. Thus the soliton solution eq.~(28) is a winding number one map. It
looks singular at $|\vec x|=0$
because the $\xi$ coordinate system for the Goldstone
boson manifold is singular at the equator of $S^3$. However, this
singularity is a reflection of the coordinate singularity on the
Goldstone boson manifold in the $\xi$ basis and is not a physically
singular field configuration.

The form of the Lagrangian density for heavy meson-pion interactions in
eq.~(22) is not very convenient for discussing heavy meson-soliton
interactions because of the coordinate singularity in $\xi$. It is convenient
when dealing
with matter fields interacting with solitons to redefine the fields so
that the Lagrangian density is expressible in terms of $\Sigma$.  For
example, introducing new heavy meson fields $P'_a$ and $P^{'*}_{a\mu}$,
defined by
$$      H'_a = H_b \xi_{ba} \,\, , \eqno (29)$$
the $SU(2)_L \times SU(2)_R$ transformation law in eq.~(18) becomes
$$      H'_a \rightarrow H'_b R^{\dagger}_{ba} \,\, , \eqno (30)$$
and the Lagrangian density in eq.~(22) becomes
$$      {\cal L} = - i Tr \bar H'_a v_\mu \partial^\mu H'_a
+ {i\over 2} Tr
\bar H'_a H'_b v^\mu (\Sigma^{\dagger} \partial_\mu \Sigma)_{ba}$$
$$      + {ig\over 2} Tr \bar H'_a H'_b \gamma^\nu \gamma^5
(\Sigma^{\dagger} \partial_\nu \Sigma)_{ba} + ... \eqno (31)$$
The parity transformation rule for the primed heavy meson fields
is a little more complicated,
$$      H'_a (x^0, \vec x) \rightarrow \gamma^0 H'_b (x^0, - \vec x)
\gamma^0 \,\Sigma^{\dagger}_{ba} (x^0, - \vec x) \,\, . \eqno (32)$$
Note that in a background Goldstone boson field configuration
of a soliton located at the same spatial
point as the heavy meson the factor of $\Sigma^{\dagger}$ becomes $-1$, whereas
$\Sigma^{\dagger}=1$ for a heavy meson infinitely far from the soliton.
This relative minus sign is the source of the parity flip that gives positive
parity heavy meson-soliton bound states.

In the large $N_c$ limit, the baryon
solitons $B$ containing light $u$ and $d$ quarks
are very heavy and time derivatives on the pion fields can be neglected.
Consequently  the interaction Hamiltonian
$$      H_I = - {ig\over 2} \int d^3 \vec x \
Tr \bar H'_a H'_b \gamma^j \gamma_5
[\Sigma^{\dagger} \partial_j \Sigma]_{ba} + ...\,\, , \eqno (33)$$
with $\Sigma$ given by eq.~(25) determines the potential energy of a
configuration with a baryon $B$ at the origin and a heavy $P'$ or $P'^{*}$
meson at position $\vec
x$.  Using eq.~(16) and taking the trace in eq.~(33) gives
$$
H_I = g \int d^3 \vec x \ [P'^{*j\dagger}_a (\vec x) P'_b (\vec x) +
P'^{\dagger}_a (\vec x) P'^{*j}_b (\vec x)
+i\epsilon^{ikj} P'^{*i\dagger}_a (\vec x) P'^{*k}_b(\vec x)]$$
$$
\times\Bigg[ A\left\{ {x^j\over |\vec x|}\hat x \cdot \tau\left(F'
- {\sin (2F)\over 2|\vec x|}\right) + {\tau^j\over 2|\vec x|} \sin
(2F)
+ \epsilon^{jmk} x^k \tau^m {\sin^2 (F)\over |\vec x|^2}
\right\} A^{-1} \Bigg]_{ba} + ...  \eqno (34)$$
Assuming that in attractive channels the potential energy is minimized
at $\vec x = \vec 0$ where the heavy meson and baryon soliton are located at
the same point, it is eigenvalues of the potential operator at the
origin that are needed to determine the spectrum of bound states.
Using eq.~(34)
$$      V_I (\vec 0) =  g F' (0) [P'^{*j\dagger}_a P'_b + P'^{\dagger}_a
P'^{*j}_b + i\epsilon^{ikj} P'^{*i\dagger}_a P'^{*k}_b]$$
$$      \times [(a_0 + i \vec a \cdot \vec \tau)\tau^j (a_0 - i \vec a
\cdot \vec \tau)]_{ba} + ... \,\, , \eqno (35)$$
where the ellipsis denotes the contribution of terms in the chiral
Lagrangian with more than one derivative.  It is easy to
diagonalize the  potential (at the origin) matrix
that arises from taking the {\it truncated} basis of nucleon-heavy meson
product states (i.e., $|p\uparrow
\rangle|P'_1\rangle$, etc.).  On this space it is
straightforward to show that\foot{More correctly
this gives the matrix elements of the potential
operator in eq.~(35) divided by the normalization of states.  The
conventions for the field operators and state normalizations are the
same as those in Ref. [14].}
$$      V_I (\vec 0) = - {2\over 3} g F' (0) (\vec S_\ell^{\,2} - 3/2) (\vec
I^{\,2} - 3/2) + ... \,\, , \eqno (36)$$
where $\vec S_\ell$ denotes the total angular momentum vector of the light
degrees of freedom (soliton and heavy meson combined)
and $\vec I$ denotes the total isospin vector.
The eigenstates of $V_I (\vec 0)$
have definite isospin, spin and angular momentum for the light degrees
of freedom.  They are denoted by $|I, s, s_\ell\rangle$,
where $I$ is the total isospin, $s$ the total spin and $s_\ell$ is the
angular momentum of the light degrees of freedom.

The spatial wavefunctions of the eigenstates of eq.~(36) are
$\delta^3\left(\vec x\right)$, and are parity even. The parity of the
meson-soliton bound state is also even, because the primed heavy meson
fields are odd under parity
at infinity, and so are even under parity at the origin, as noted below
eq.~(32). The unprimed heavy meson
fields have a simple transformation law under parity, and
do not have a relative minus sign between the
parity at infinity and parity at the origin. However, in the $\xi$
basis, the wavefunction of the bound state contains a factor of the
form $\vec\tau\cdot\hat x$ near the origin.
Thus the parity
of the soliton is also even in the $\xi$ basis because the negative
parity of the spatial wavefunction is combined with the negative
intrinsic parity of the heavy meson field. The factor $\vec\tau
\cdot\hat x$ appears singular at the origin, but that is because of the
coordinate singularity in the $\xi$ basis. Any physical quantity
is well-defined at the origin.

It is interesting to see if the interaction in eq.~(33) gives a
reasonable
spectrum of baryons containing a heavy quark when the effects of
operators with more than one derivative
are neglected. The first column of energies in Table 1 gives the eigenstates
and eigenvalues of  $V_I(\vec
0)$ in the truncated basis.  Only the $|0, 1/2, 0\rangle, |1, 1/2, 1\rangle$
and $|1,
3/2, 1\rangle$
states are bound if $g$ is taken positive.  For the case $Q = c$ these
states have the right $I,s, s_\ell$ and parity quantum numbers to be the
$\Lambda_c, \Sigma_c$ and $\Sigma_c^*$ respectively.
In the large $N_c$ limit, the $N$ and $\Delta$ are degenerate and the space of
states should be enlarged to include products of $\Delta$-baryons with the $P'$
and $P^{*'}$ mesons. Such states can have the same quantum numbers as the
$\Sigma_c$ and the $\Sigma_c^*$. Because the spin-spin interaction is
suppressed by $1/N_c$, including the $\Delta$ causes the $\Sigma_c$ and
$\Sigma_c^*$ to be degenerate with the $\Lambda_c$ and gives the energies
presented in the last column of Table 1.${}^{19}$
The real world is intermediate between the $N_c\rightarrow\infty$ limit with a
degenerate $N$ and $\Delta$, and the limit where the $\Delta$ is neglected
because it is considered to be much heavier than the nucleon. Including the
$\Delta$ with a finite $\Delta-N$ mass difference will lead to a $\Sigma_c$
state which is heavier than the $\Lambda_c$ with a binding energy that is
intermediate between the values in the two columns of Table~1. Thus the
interaction in eq.~(33) gives a reasonable qualitative description of heavy
baryons. However it is important to remember that anti-baryon-meson bound
states are described by taking $F\rightarrow -F$ so the interaction in eq.~(33)
gives rise to exotic states. These states can be removed by including the
effects of higher derivative terms in the chiral lagrangian.
This is not neccessarily in conflict with the derivative expansion, since the
normal state binding energy is three times the exotic state binding energy.

In the $N_c\rightarrow\infty$ and
$m_Q\rightarrow\infty$ limit, the most general
potential at the origin including all
possible higher dimension operators
has the form
$$
V = V_0(F) -
{2\over 3} V_1(F) (\vec S_\ell^{\,2} - 3/2) (\vec
I^{\,2} - 3/2) \,\, , \eqno (39)$$
on the truncated meson-nucleon subspace. In eq.~(39)
$V_0$ and $V_1$ are functionals of the soliton shape function $F(|\vec x|)$.
The coefficients $V_0$ and $V_1$ have no definite symmetry under $F\rightarrow
-F$, so that the  meson-nucleon and meson-antinucleon scattering potentials are
not related to each other.

The work presented in this paper is similar in spirit to that of Callan
and Klebanov.$^{10}$  The main difference is that constraints imposed by
heavy quark symmetry are taken into account.  The bound state approach
of Callan and Klebanov cannot be taken over to baryons containing a
heavy charm or bottom quark without imposing heavy quark symmetry
(e.g., there is no reason
to expect eq.~(6.2) of Ref. [10] to hold).  The mechanism for the parity
flip presented in this paper is the same as that of Callan
and Klebanov.  For example, if the kinetic energy of the nucleon is
included then the ground state is an $L = 0$ partial wave,\foot{
With a radial wave function that is strongly peaked about
$|\vec x| = 0$.}
instead of a Dirac delta function.  (Some of the excited negative parity
baryons containing a heavy quark appear as bound states with odd orbital
angular momentum.)  However, since the
unprimed and primed heavy meson fields are related (near the origin) by
a factor of $\hat x \cdot \vec \tau$ the $S$-wave primed heavy
meson-nucleon bound state wave functions get multiplied by a linear
combination of $Y_{1,m}$ spherical harmonics in the unprimed basis and
hence appear as $P$-waves.  Similar changes in quantum numbers also
occur in bound states involving monopoles.$^{20}$

\medskip
Mark B. Wise thanks N. Seiberg for suggesting that it would be interesting to
examine how baryons containing a heavy quark arise as solitons.

\bigskip
\vskip2truecm

\input tables
\centerline{{\bf Table 1}}
\bigskip
\begintable
States | ~~Energies neglecting the $\Delta$~~ | ~~Energies including the
$\Delta$~~\cr
$\vert 0, 1/2, 0\rangle$ | $- 3 g F'(0)/2$ | $-3 g F'(0)/2$\crnorule
$\vert 1, 3/2,1\rangle , \vert 1, 1/2, 1\rangle$ | $- g F'(0)/6$ | $-3 g
F'(0)/2$
\crnorule
$\vert 1, 1/2, 0\rangle$ | $g F'(0)/2$ | $g F'(0)/2$\crnorule
$\vert 0, 3/2, 1\rangle, \vert0, 1/2, 1\rangle$ | $ g F'(0)/2$ | $g F'(0)/2
$\endtable
\vfil\eject

\noindent {\bf References}

\item{1.}  E. Witten, Nucl. Phys. {\bf B160}, 57 (1979).

\item{2.}  T.H.R. Skyrme, Proc. Roy. Soc. {\bf A260}, 127 (1961).

\item{3.}  E. Witten, Nucl. Phys. {\bf B223}, 433 (1983).

\item{4.}  J. Wess and B. Zumino, Phys. Lett. {\bf B37}, 95 (1971).

\item{5.}  G.S. Adkins, C.R. Nappi and E. Witten, Nucl. Phys. {\bf
B228}, 552 (1983).

\item{6.}  M. Mattis and M.E. Peskin, Phys. Rev. {\bf D32}, 58 (1985);
M. Karliner and M. Mattis, Phys. Rev. {\bf D34}, 1991, (1986).

\item{7.}  N. Isgur and M.B. Wise, Phys. Lett. {\bf B232}, 113 (1989);
Phys. Lett. {\bf B237}, 527 (1990).

\item{8.}  H. Georgi, Phys. Lett. {\bf B240}, 447 (1990).

\item{9.}  N. Isgur and M.B. Wise, Phys. Rev. Lett. {\bf 66}, 1130
(1991).

\item{10.}  C.G. Callan and I. Klebanov, Nucl. Phys. {\bf B262}, 365
(1985), Phys. Lett. {\bf B202}, 269 (1988).

\item{11.} M. Rho, D.O. Riska, and N.N. Scoccola, Phys. Lett. {\bf B
251}, 597, (1990);
Y. Oh, D. Min, M. Rho, and N. Scoccola, Nucl. Phys. {\bf A534}, 493
(1991); M. Rho, D.O. Riska, and N.N. Scoccola, Z. Phys. {\bf A341},
343, (1992).

\item{12.} S.~Coleman, J.~Wess, and B.~Zumino, Phys. Rev. {\bf 177},
2239 (1969); C.~Callan, S.~Coleman, J.~Wess, and B.~Zumino, Phys. Rev.
{\bf 177}, 2247 (1969).

\item{13.}  A.F. Falk, H. Georgi, B. Grinstein and M.B. Wise, Nucl.
Phys. {\bf B343}, 1 (1990); J.D. Bjorken, SLAC-PUB-5278, invited talk at
Les Rencontres de Physique de la Vallee d'Aoste, La Thuile, Italy (1990)
unpublished; H. Georgi, HUTP-91-A039, lectures delivered at TASI, (1991)
unpublished.

\item{14.}  M. Wise, Phys. Rev. {\bf D45}, 2118 (1992).

\item{15.}  G. Burdman and J.F. Donoghue, Phys. Lett. {\bf B280}, 287 (1992).

\item{16.}  Tung-Mow Yan, Hai-Yang Cheng, Chi-Yee Cheung, Gueg-Lin Lin,
Y.C. Lin and Hoi-Lai Yu, CLNS 92/1138 (1992) unpublished.

\item{17.}  The ACCMOR Collaboration (S. Barlag, et al.), Phys. Lett. {\bf
B278}, 480 (1992).

\item{18.} H. Georgi, {\it Weak Interactions and Modern Particle
Theory\/}, (Benjamin/Cummings, 1984);
A. Manohar and H. Georgi, Nucl. Phys. {\bf B234}, 189 (1984).

\item{19.} Z. Guralnik, M. Luke and A.V. Manohar, UCSD/PTH 92-24 (1992).

\item{20.}  See, for example, S. Coleman, The Magnetic Monopole Fifty Years
Later, in {\it The Unity of Fundamental Interactions\/}, ed. by A. Zichichi,
(Plenum, New York, 1983).

\bye